\documentclass[journal]{IEEEtran}

\ifCLASSINFOpdf

\else

\fi

\usepackage{enumitem}
\usepackage{pdfcomment}
\usepackage{amsmath}
\usepackage{amsfonts}
\usepackage{longtable}
\usepackage{graphicx}
\usepackage{ragged2e} 
\usepackage{xurl}
\usepackage{hyperref}
\usepackage{cite}
\usepackage{xcolor}
\usepackage{etoolbox}
\usepackage{xspace}
\usepackage{url}
\usepackage{microtype}
\usepackage[compatibility=false]{caption}
\usepackage{subcaption} 

\begin{document}
	
	\title{Explainability-Aware Evaluation of Transfer Learning Models for IoT DDoS Detection Under Resource Constraints}
	
	\author{Nelly Elsayed, ~\IEEEmembership{Senior,~IEEE}}


	%



	\maketitle
	
	\begin{abstract}
		Distributed denial-of-service (DDoS) attacks threaten the availability of Internet of Things (IoT) infrastructures, especially in resource-constrained deployments. Although transfer learning models have shown promising detection accuracy, their reliability, computational feasibility, and interpretability in operational environments remain insufficiently explored. This study presents an explainability-aware empirical evaluation of seven pre-trained convolutional neural network architectures for multi-class IoT DDoS detection using the CICDDoS2019 dataset and an image-based traffic representation.
		The analysis integrates classification performance metrics, complementary statistical indicators (MCC, Cohen's kappa, and Youden Index), training and inference cost assessment, and interpretability evaluation using Grad-CAM and SHAP.
		Results indicate that DenseNet and MobileNet-based architectures achieve strong detection performance while demonstrating superior reliability and compact, class-consistent attribution patterns.
		MobileNet achieves the highest reported classification accuracy and several of the strongest reliability statistics, whereas DenseNet169 exhibits the most favorable qualitative interpretability characteristics based on Grad-CAM and SHAP. MobileNetV3 offers a potential trade-off between inference latency, classification performance, and qualitative interpretability for resource-constrained deployment. The findings emphasize the importance of combining performance, reliability, and explainability criteria when selecting deep learning models for IoT DDoS detection.
	\end{abstract}
	

	%
	\IEEEpeerreviewmaketitle

	\section{Introduction}

	The Internet of Things (IoT) has emerged as a foundational component of modern digital infrastructures~\cite{ebrahimpour2024authentication,alaba2017internet}. It has been widely deployed across large-scale domains, including industrial systems~\cite{kotsiopoulos2025defending,behnke2023real,reyes2025analysis}, healthcare infrastructures~\cite{selvaraj2020challenges,elsayed2023machine,azzawi2016review}, and enterprise networks~\cite{jayashree2020introduction}, as well as in smaller-scale environments such as smart homes~\cite{wijesundara2024security,samuel2016review,alaa2017review}, small businesses~\cite{jones2018can,jeong2017study}, and office settings~\cite{shetty2024iot}. While IoT enables automation and real-time monitoring, its rapid expansion also significantly increases exposure to cyber threats~\cite{shaheen2025systematic,nappi2020internet}. Integrating resource-constrained devices, heterogeneous communication protocols, and distributed architectures introduces substantial security vulnerabilities that can affect hardware assets, sensitive data, and, most critically, system availability~\cite{sun2025survey}. Therefore, securing IoT infrastructure and ensuring resilience are primary cybersecurity concerns.
	
\begin{figure}[t]
		\centering
		\pdftooltip{
			\includegraphics[width = 5.5cm, height=5cm]{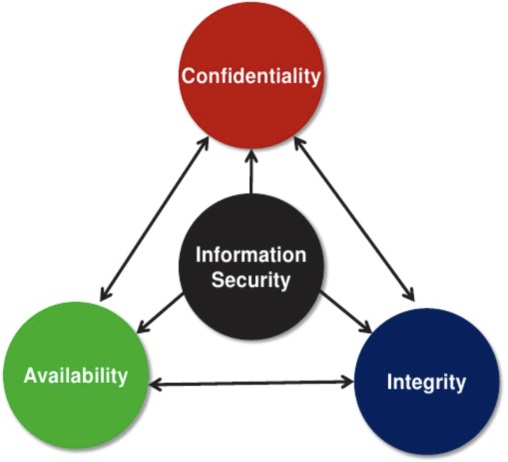}}{The CIA triad consists of Confidentiality, Integrity, and Availability as the angles of a triangle, and at the center of the triangle, information security connects all three.}
		\caption{The cybersecurity CIA triad represents the three pillars of information security: confidentiality, integrity, and availability~\cite{nist_sp1800_26a}.}
		\label{triad}
	\end{figure}

	Denial of Service (DoS) and Distributed Denial of Service (DDoS) attacks are among the most significant threats to IoT-driven environments~\cite{vishwakarma2020survey,dzamesi2025review}. These attacks aim to exhaust bandwidth, computational capacity, or protocol state resources by generating overwhelming traffic volumes~\cite{cisco1,elsayed2025cryptodna}. This resource depletion can cause partial or complete service disruption for legitimate users~\cite{eliyan2021and,dvzaferovic2019and}. From a cybersecurity perspective, such attacks directly violate the availability principle of the information security pillars: Confidentiality–Integrity–Availability (CIA) triad~\cite{nist_sp1800_26a}, as shown in Figure~\ref{triad}, which represents the foundational framework for information security~\cite{samonas2014cia}. Therefore, preserving availability in IoT infrastructures is essential to maintaining operational continuity and preventing large-scale service outages.

	Recent reports indicate a substantial increase in DDoS activity worldwide. According to VERCARA~\cite{vercara2024}, the industry analysis report shows a significant year-over-year growth in volumetric and reflection-based DDoS campaigns targeting telecommunications, finance, and service providers. IoT environments are particularly vulnerable due to limited computational resources, constrained memory, and often insufficient built-in security mechanisms~\cite{kumari2023comprehensive,yan2025intelligent}. 
	
	DDoS attacks against IoT networks take multiple forms, including volumetric flooding attacks, protocol exploitation attacks, and application-layer attacks~\cite{kumari2023comprehensive,al2021ddos,sharif2023detection}. Reflection and amplification techniques, leveraging services such as DNS, NTP, SSDP, and LDAP, enable attackers to magnify traffic volume while spoofing source addresses. Attackers commonly execute such attacks via botnets comprising compromised devices distributed across multiple regions~\cite{alomari2012botnet,elsayed2023iot,pandey2025conditional}. The distributed nature of botnets makes detection and mitigation particularly challenging~\cite{hoque2015botnet,brezo2011challenges,al2025deep}.
	
	Traditional detection mechanisms, such as firewalls, rule-based intrusion detection systems (IDS), and manual traffic monitoring, often struggle to maintain performance under complex, multi-vector DDoS attacks~\cite{sharafaldin2019developing,khalaf2019comprehensive,ghazanfar2020iot}. Thus, researchers have increasingly adopted machine learning and deep learning approaches to identify anomalous traffic patterns~\cite{doshi2018machine,zaghloul2021green,almobaideen2025comprehensive,elsayed2025extreme,dougan2025detection,kaur2023investigation,elsayed2025and,mittal2023deep,hizal2024novel,roopak2020intrusion,agrawal2024detection}. However, several practical challenges remain insufficiently addressed. Many existing studies prioritize detection accuracy while overlooking computational cost, inference latency, deployment feasibility in resource-constrained IoT environments, and the reliability and transparency of model decisions~\cite{quincozes2024survey,elsayed2022autonomous}.

	In this paper, we aim to address these challenges through a security-grounded empirical evaluation of pre-trained deep learning models for DDoS detection in IoT infrastructures. Specifically, this paper investigates the following core questions: 
	\begin{enumerate}
		\item How effectively can pre-trained models detect heterogeneous DDoS attack types in IoT environments? 
		\item What are the computational and latency trade-offs associated with deploying such models under fog-level resource constraints? 
		\item To what extent can explainable artificial intelligence techniques enhance the reliability and transparency of detection decisions?
	\end{enumerate}
	
	To address these questions, we evaluate multiple lightweight and advanced pre-trained models using a benchmark IoT DDoS dataset. We analyze the detection performance, computational overhead, and inference time, and integrate SHAP and Grad-CAM explainability methods to assess decision reliability~\cite{alenezi2021explainability,tempel2025choose}. By combining performance evaluation with interpretability analysis, this paper seeks to inform the selection of reliable and transparent DDoS detection models for resource-constrained IoT infrastructures.

	
	\section{Threat Model and Security Assumptions}
	
	In this paper, we consider the distributed denial-of-service (DDoS) threat targeting IoT-driven network infrastructures. The threat model formally characterizes adversarial capabilities, defender assumptions, and the security objectives guiding the proposed detection framework~\cite{Tatam_Shanmugam_Azam_Kannoorpatti_2021,Apruzzese2022}.

	\subsection{Adversary Model}
	
	We assume a remote adversary controlling a botnet composed of compromised devices capable of generating high-volume traffic toward a target IoT infrastructure. The attacker can launch volumetric, protocol-based, and application-layer DDoS attacks, including reflection and amplification attacks such as DNS, NTP, SSDP, LDAP, MSSQL, UDP floods, SYN floods, and related variants~\cite{bottger2015dos,antonakakis2017mirai,rossow2014amplification,mirkovic2004taxonomy,durumeric2015searching,kolias2017ddos}. The adversary is capable of generating large-scale distributed traffic from multiple sources, spoofing source IP addresses in reflection-based attacks, and varying attack strategies across known DDoS categories to exhaust network and computational resources.
	
	However, we do not assume the adversary performs adversarial machine learning attacks such as model poisoning during training or evasion through carefully crafted perturbations targeting model decision boundaries~\cite{biggio2018wild,Giattino_2019}. Furthermore, we assume that the attacker does not gain direct access to modify or tamper with the deployed detection model. This paper focuses on network-level DDoS detection rather than adversarial ML robustness.
	
	\subsection{Defender Model}
	
	The defender operates at the fog or gateway level within the IoT infrastructure, where it monitors and detects traffic before it reaches core services~\cite{stojmenovic2014fog,yi2015survey,4738466}. The detection system monitors network flow-level features extracted from traffic without performing deep packet inspection or payload analysis~\cite{moore2005internet}. The model is trained offline using labeled data and deployed for real-time inference to support timely attack identification~\cite{GARCIA2014100,5504793}.
	
	The defender's objectives are to detect DDoS attacks while reliably preserving service availability. In particular, minimizing false negatives is critical to prevent prolonged service disruption. Controlling false positives is essential to reduce unnecessary mitigation actions and operational overhead~\cite{axelsson2000intrusion,5356528}.

	\subsection{Security Objectives}
	
	Our proposed framework's main security goal is to preserve the availability of IoT-driven systems under DDoS conditions. To achieve this objective, the detection mechanism aims to identify attack traffic quickly and accurately, maintain stability across heterogeneous attack categories, and support security assurance through explainable decision-making. The framework seeks to enhance transparency and trust in automated detection within resource-constrained IoT environments
	by integrating performance evaluation with interpretability analysis.
	
	\section{Background and Related Work}
	
	\subsection{IoT Architecture and Attack Surface}
	
\begin{figure*}[htbp]
		\centering
		\pdftooltip{
			\includegraphics[width=8cm,height=9cm]{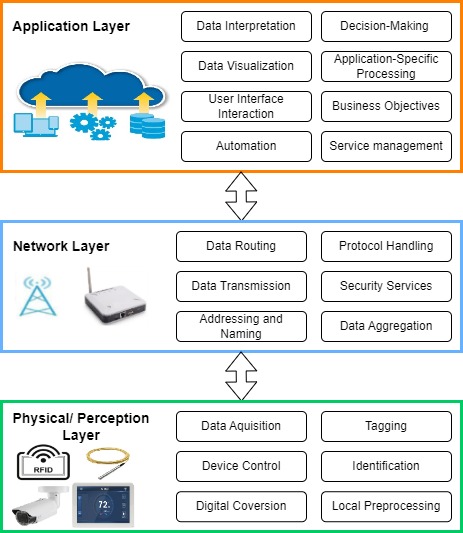}
		}{The IoT three-layered architecture consists of the Application Layer, the Network Layer, and the Physical/Perception Layer.}
		\caption{The IoT three-layered architecture~\cite{azumah2021deep}.}
		\label{iotLayers}
	\end{figure*}

	Figure~\ref{iotLayers} illustrates the three logical layers of IoT architecture: the perception layer, the network layer, and the application layer~\cite{alasmary2024scalabledigitalhealth}. The perception layer comprises sensors and embedded devices that acquire data. The network layer enables communication between IoT devices, gateways, and external services through heterogeneous wired and wireless protocols. The application layer delivers monitoring, analytics, and control services~\cite{jabraeil2020iot,olivier2015new}.
	
	The distributed and heterogeneous nature of IoT systems increases the attack surface, especially at the network layer, where traffic aggregation and routing occur. Limited computational capacity, constrained memory, and inconsistent security configurations further amplify vulnerability to availability-oriented attacks~\cite{atlam2020iot}. As IoT deployments scale across industrial and enterprise environments, protecting network availability becomes a central security requirement.
	
	\subsection{DDoS Threats in IoT Environments}
	
	Distributed denial-of-service (DDoS) attacks are considered among the most prevalent threats to IoT infrastructures~\cite{kumari2023comprehensive,Li_Wei_Wu_Guo_2023,conti2018survey}. DDoS attacks aim to exhaust bandwidth, computational resources, or protocol state tables by generating high volumes of malicious traffic~\cite{Mirko2005,specht2004taxonomy}. In IoT environments, attackers frequently exploit reflection and amplification mechanisms such as DNS, NTP, SSDP, and LDAP to magnify traffic volume while spoofing source addresses~\cite{czyz2014measuring}.
	
	Botnets composed of compromised IoT devices enable geographically distributed traffic generation at scale~\cite{8638982,elsayed2023iot,pandey2025conditional}. Due to limited built-in security mechanisms and irregular patch management, IoT devices often serve both as attack targets and as components of attack infrastructure~\cite{stellios2018survey}. Consequently, reliable and efficient DDoS detection mechanisms are essential to preserve service availability in IoT-driven systems.
	
	\subsection{Machine Learning-Based DDoS Detection}
	
	Traditional rule-based intrusion detection systems and signature-based mechanisms often struggle to cope with large-scale, multi-vector DDoS attacks~\cite{almobaideen2025comprehensive,sharafaldin2019developing}. Thus, researchers have widely employed machine learning (ML) and deep learning (DL) techniques for traffic classification and anomaly detection in IoT environments to overcome these limitations. 
	
	Several prior studies explored classical ML algorithms, including support vector machines, random forests, extreme learning machines, and k-nearest neighbors, for DDoS detection using handcrafted traffic features. More recently, deep neural networks, convolutional neural networks (CNNs), and recurrent architectures have demonstrated improved detection accuracy by learning complex traffic patterns directly from data~\cite{almobaideen2025comprehensive,bala2024ai,elsayed2025extreme,kaur2021comprehensive,dougan2025detection}. Transfer learning approaches using pre-trained CNN architectures have also been investigated to reduce training cost and leverage feature extraction capabilities developed in large-scale image classification tasks~\cite{goodfellow2016deep,kaur2023investigation,sanmorino2024fine}.
	
	Despite promising detection performance, several limitations remain in the existing literature. Several research studies emphasize classification accuracy while providing limited analysis of computational cost, inference latency, and deployment feasibility in resource-constrained IoT or fog-level environments~\cite{sanmorino2024fine,sanmorino2024detection,kachavimath2025detection}. In addition, reliability evaluation often relies on aggregate performance metrics and does not examine stability across heterogeneous attack types~\cite{sharafaldin2019developing}. Moreover, the explainability and interpretability of deep learning-based detection models still raise concerns regarding transparency, trustworthiness, and forensic usability in operational settings~\cite{mohale2025systematic,prasad2026explainable}.
	
	This paper addresses these limitations through a comparative evaluation of seven pretrained deep learning architectures for IoT DDoS detection. The evaluation combines classification performance, computational cost, statistical reliability indicators, and post-hoc explainability analyses to inform model selection under resource constraints.
	
	\section{Research Methodology}
	
	In this section, we present the research components and methodology of this empirical and analytical study to answer the core research questions. Our main objective is to systematically evaluate pre-trained deep learning models for detecting DDoS attacks in IoT-driven environments under fair comparison constraints.
	
	To ensure transparency, reproducibility, and fair comparisons among models, we used a publicly available, standardized benchmark dataset for DDoS attacks, allowing the research community to replicate and extend the study in future investigations.
	
\begin{figure*}[htbp]
		\centering
		\pdftooltip{
			\includegraphics[width=14cm,height=5.7cm]{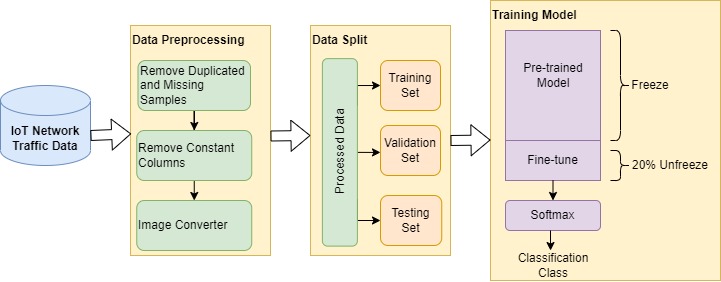}}{Flowchart illustrating the IoT DDoS detection methodology. IoT network traffic data undergoes preprocessing to remove duplicate and missing samples, eliminate constant columns, and convert the data into images. The processed data is divided into training, validation, and testing sets. The model then fine-tunes a pre-trained model by unfreezing 20\% of its layers while keeping the rest frozen. A softmax classifier produces the final traffic classification.}
		\caption{The proposed research methodology for investigating pre-trained models for DDoS attack detection.} 
		\label{oveall_methodology}
	\end{figure*}

	Figure~\ref{oveall_methodology} shows the overall research methodology used to investigate pre-trained models for DDoS attack detection.
	
	\subsection{Dataset}
	
	In this paper, we used the CICDDoS2019 benchmark dataset~\cite{sharafaldin2019developing}, which is publicly available and widely adopted in DDoS detection research. The dataset contains 12 labeled network traffic classes representing benign traffic and multiple DDoS attack categories~\cite{somani2017ddos,bala2024ai}. The considered attack types include DNS amplification, LDAP amplification, MSSQL reflection, NetBIOS, NTP amplification, SNMP amplification, SSDP, UDP flood, SYN flood, TFTP amplification, and UDP-Lag attacks.
	
	Each attack category represents distinct traffic characteristics, including volumetric flooding, protocol exploitation, and reflection-based amplification mechanisms. This diversity enables the evaluation of model robustness across heterogeneous attack patterns.

	\begin{figure}[t]
		\centering
		\pdftooltip{
			\includegraphics[width=5cm,height=3cm]{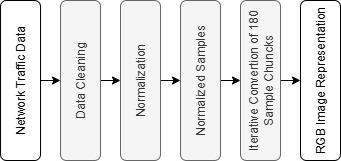}}{Flowchart illustrating how network traffic data is converted into RGB image representations. The process consists of six sequential steps: network traffic data acquisition, data cleaning, normalization, preparation of normalized samples, iterative conversion into chunks of 180 samples, and generation of RGB image representations for deep learning-based DDoS detection.}
		\caption{Network traffic data transformation into image representation for pre-trained models.}
		\label{Data_Transformation}
	\end{figure}
	
	We preprocessed the dataset by removing duplicate records, incomplete samples, and constant-valued features to reduce redundancy and eliminate non-informative attributes. We scaled features using min–max normalization to ensure consistent value ranges across all features. The normalization process is defined as:
	
	\begin{equation}
		X' = \frac{X - \min(X)}{\max(X) - \min(X)} \times 255
	\end{equation}
	
	where \(X\) represents the original feature value and \(X'\) denotes the normalized value scaled to the [0–255] range suitable for image representation.

	We transformed structured tabular traffic data into image-based representations following established approaches~\cite{liu2019predicting,liu2019novel,li2017intrusion}, specifically adopting the traffic-to-image conversion procedure described by Hussain et al.~\cite{hussain2020iot}. In this procedure, data cleaning and feature selection retain 60 network-flow features per record. Following feature-wise min--max normalization to the range $[0,255]$, groups of 180 traffic records are arranged into three-channel images of size $60 \times 60 \times 3$, with 60 records assigned to each RGB channel. This representation enables the application of pretrained convolutional neural networks to network-traffic classification.

	\subsection{Pre-trained Models Selection}
	
	The pre-trained models were selected based on the following criteria:
	
	\begin{itemize}
		\item The model is designed for image classification within transfer learning frameworks.
		\item The model is available in TensorFlow/Keras for reproducibility and deployment compatibility.
		\item The model can process the transformed traffic images directly or through input resizing compatible with the selected architecture (less than a 20\% upscale of the original image size).
		\item The total number of parameters is below 15 million parameters to align with resource constraints typical in IoT and fog-level deployments.
	\end{itemize}
	
	We selected TensorFlow/Keras to support deployment in edge and mobile environments because of its wide industry adoption and support for lightweight inference engines.
	
	The parameter counts in Table~\ref{pre_trained_models} are approximate reference values for the pretrained architectures. Table~\ref{compare_models} reports the trainable and non-trainable parameter counts recorded for the task-adapted models used in this study.
	
	\begin{table*}[htbp]
		\centering
		\renewcommand{\arraystretch}{1.3}
		\caption{Comparison of selected pre-trained models based on selection criteria.}
		\label{pre_trained_models}
		\begin{tabular}{lllll}
			\hline
			\textbf{Model} & \textbf{Model Description} & \textbf{No. Param.} & \textbf{Type} & \textbf{Handle Input Size} \\
			MobileNet & Lightweight CNN for mobile devices & 4M & Lightweight & Yes \\
			MobileNetV2 & Improved MobileNet with depthwise separable convolutions & 3.4M & Lightweight & Yes \\
			MobileNetV3 & Optimized MobileNet version for efficiency & 5.4M & Lightweight & Yes \\
			DenseNet121 & Densely connected CNN with fewer parameters & 8M & CNN-Based & Yes \\
			DenseNet169 & Deeper DenseNet variant & 14M & CNN-Based & No \\
			EfficientNet-B0 & Optimized for accuracy–efficiency balance & 5.3M & Advanced Architecture & No \\
			EfficientNetB1 & Larger EfficientNet variant with higher capacity & 7.8M & Advanced Architecture & No \\
			\hline
		\end{tabular}
	\end{table*}
	
	\subsection{Implementation Specification}
	
	The optimization strategy plays a significant role in ensuring stable convergence during fine-tuning. In this study, we used Stochastic Gradient Descent (SGD) because of its stability and controlled parameter updates in image classification tasks~\cite{klein2009adaptive,dogo2018comparative}. 
	
	We applied transfer learning with partial fine-tuning by unfreezing the top layers of each model. Fine-tuning requires careful learning-rate control. The reference configuration used an initial learning rate of \(lr = 0.001\) and a ReduceLROnPlateau scheduler. 
	
	We evaluated all models within the same benchmark-based experimental framework. Any architecture-specific training configurations are relevant when interpreting differences in results. We evaluated performance on the independent test set using standard classification metrics.
	
	\subsection{Experimental Setup}
	
	We implemented all experiments in Python using TensorFlow/Keras. We stored and loaded the transformed traffic samples in HDF5 format using the \texttt{h5py} library. We normalized each input image before training by scaling pixel values to the $\mathrm{[0,1]}$ range by dividing by 255. We encoded class labels using one-hot encoding to support multi-class classification across benign traffic and the DDoS categories included in CICDDoS2019.

	\subsubsection{Data Splitting and Evaluation Protocol}
	
	To ensure a fair comparison across architectures, we trained and evaluated all models using the same data-splitting protocol. We split the dataset into three sets with a ratio of 60\% for training, 20\% for validation, and 20\% for testing. We used stratified sampling to preserve the class distribution proportionally among the three sets.
	
	\subsubsection{Transfer Learning and Fine-tuning Strategy}
	
	We employed a transfer learning approach using ImageNet-pretrained convolutional neural networks as feature extractors. We then fine-tuned the model to adapt the representations to the domain of network traffic images. We excluded the original classification layers by setting \texttt{include\_top=False} and added a new classification head. The new classification head consists of a global average pooling layer, a fully connected layer with Rectified Linear Unit (ReLU) activation, and a softmax output layer corresponding to the number of traffic classes.
	
	The intended architecture-independent fine-tuning policy was to freeze approximately the first 80\% of each ImageNet-pretrained backbone and unfreeze its final 20\% of layers for domain adaptation. This percentage refers to the \emph{number of backbone layers}, not to the number or proportion of trainable parameters: the latter varies with layer type, architecture, and the task-specific classification head. Table~\ref{compare_models} records the reported layer counts and parameter partitions. 
	
	\subsubsection{Training Configuration}
	
	All models were trained using categorical cross-entropy loss and the Stochastic Gradient Descent (SGD) optimizer with momentum. We used a batch size of 32 and a ReduceLROnPlateau learning-rate scheduler. 
	
	\subsubsection{Cost and Timing Measurement}
	
	To quantify computational cost and fog-level feasibility, we recorded both total training time and per-sample inference latency. We measured training time as the elapsed time to complete full training for each model, and we computed inference latency as the average prediction time per test sample. These measurements characterize runtime cost on the experimental hardware and are reported in Table~\ref{tab:time_cost}.

	\subsubsection{Performance and Reliability Measurement}
	
	We evaluated classification performance using accuracy, precision, recall, F1-score, and the area under the receiver operating characteristic curve (AUC). Cohen's kappa was used to measure agreement between predicted and true labels beyond chance. To complement these performance indicators, we examined the Matthews correlation coefficient (MCC) and Youden Index. These measures provide complementary perspectives on classification agreement and sensitivity--specificity balance under multiclass conditions. The results are reported in Tables~\ref{results_compare} and~\ref{results_compare_statistics}.

	\subsubsection{Explainability and Reliability Support}
	We used Grad-CAM to examine spatial activation patterns associated with model predictions and SHAP to analyze the magnitude and direction of feature attributions. The resulting visualizations were compared across architectures and traffic classes to assess qualitative differences in attribution compactness, class consistency, and background suppression. These analyses complement the classification and statistical results by providing additional insight into the spatial evidence associated with model decisions.
	
	\section{Results and Evaluations}

	\subsection{Cost Evaluation}
	For cost evaluation, we focus on two measured timing indicators: (i) elapsed model-training time and (ii) average test-time inference latency per sample. These quantities inform computational feasibility.

	\begin{table*}[htbp]
		\centering
		\renewcommand{\arraystretch}{1.3}
		\caption{Fine-tuning configurations and parameter counts of the evaluated architectures.}
		\label{compare_models}
		\begin{tabular}{p{2.6cm}p{2cm}p{3cm}p{3cm}p{3cm}}
			\hline
			\textbf{Model} &  \textbf{No. Layers} & \textbf{20\% Unfreeze} & \textbf{Trainable Param.} & \textbf{Non-trainable Param.} \\
			\hline
			MobileNet & 89 & 17 & 1,924,404 & 1,371,840 \\
			MobileNetV2 & 157 & 33 &   1,609,164& 731,584\\
			MobileNetV3 &  279&58 &   2,907,564&  1,401,632 \\
			DenseNet121 &  430& 88 & 1,785,548& 5,318,336\\
			DenseNet169 & 598 &122 &3,780,428  & 8,969,792\\
			EfficientNet-B0 & 240 & 50& 2,388,060 & 1,744,275\\
			EfficientNetB1 & 342 &70 & 4,229,676 & 2,428,327\\
			\hline
		\end{tabular}
	\end{table*}
	
	We compared the models' training time and average inference time per test sample as indicators of computational cost and runtime efficiency. Table~\ref{tab:time_cost} shows the total training time (minutes) and average test time per sample (seconds).

	\begin{table*}[htbp]
		\centering
		\renewcommand{\arraystretch}{1.3}
		\caption{Training time and average inference latency of the evaluated pretrained models.}
		\label{tab:time_cost}
		\begin{tabular}{p{2.1cm}ll}
			\hline
			\textbf{Model} & \textbf{Train Time (m)} & \textbf{Avg. test time/sample (s)}\\
			\hline
			MobileNet & 61.745 & 0.000530   \\
			MobileNetV2 & 109.633 & 0.000864 \\
			MobileNetV3 &    140.915 & 0.001093 \\
			
			
			DenseNet121 & 248.882&  0.002191   \\
			DenseNet169 & 345.697& 0.003050  \\
			EfficientNet-B0 &  183.157&  0.001427 \\
			EfficientNetB1 &   240.622&  0.001972 \\
			\hline
		\end{tabular}
	\end{table*}

	\subsection{Performance Evaluation}
	In our empirical study, we evaluated model performance using accuracy, precision, recall, F1-score, area under the receiver operating characteristic curve (AUC), and Cohen's kappa coefficient ($\kappa$), as reported in Table~\ref{results_compare}. Cohen's kappa measures agreement between predicted and true labels beyond chance and is calculated as follows:
	
	\begin{equation}
		\kappa = \frac{P_o - P_e}{1 - P_e}
	\end{equation}
	\noindent
	where $P_o$ is the observed agreement (overall accuracy), and $P_e$ is the expected agreement by chance. For multi-class classification:
	\begin{equation}
		P_o = \sum_{i=1}^{K} \frac{n_{ii}}{N}
	\end{equation}
	\begin{equation}
		P_e = \sum_{i=1}^{K} \left( \frac{n_{i\cdot}}{N} \times \frac{n_{\cdot i}}{N} \right)
	\end{equation}
	where $K$ is the number of classes, $N$ is the total number of samples, $n_{i\cdot}$ is the number of actual samples in class $i$, $n_{\cdot i}$ is the number of samples predicted as class $i$, and $n_{ii}$ is the number of correctly classified samples in class $i$.
	
	\begin{table*}[htbp]
		\centering
		\renewcommand{\arraystretch}{1.3}
		\caption{Performance comparison of pretrained models for multiclass DDoS attack detection.}
		\label{results_compare}
		\begin{tabular}{p{2.1cm}llllll}
			\hline
			\textbf{Model} & \textbf{Accuracy} & \textbf{Precision} & \textbf{Recall} & \textbf{F1-Score} & \textbf{AUC (ROC)} & \textbf{Kappa} \\
			\hline
			MobileNet & 98.476\% & 0.90895 & 0.9008 & 0.90434 & 0.94621 & 0.89711 \\
			MobileNetV2 & 98.43\% & 0.9058 & 0.9058 & 0.89591 & 0.94364 & 0.89399 \\
			MobileNetV3 & 98.059\% & 0.89205 & 0.87209 & 0.88353 & 0.93506 & 0.86870 \\
			DenseNet121 & 98.25\% & 0.89142 & 0.88689 & 0.88814 & 0.93863 & 0.88187 \\
			DenseNet169 & 98.298\% & 0.89788 & 0.89007 & 0.89788 & 0.94036 & 0.88509 \\
			EfficientNet-B0 & 93.156\% & 0.58938 & 0.59623 & 0.58938 & 0.77878 & 0.53711 \\
			EfficientNetB1 & 85.998\% & 0.1599 & 0.21485 & 0.17463 & 0.56816 & 0.05826 \\
			\hline
		\end{tabular}
	\end{table*}

	\begin{figure*}[htbp]
		\centering
		
		\begin{subfigure}[b]{0.48\textwidth}
			\centering
			\pdftooltip{
				\includegraphics[width=\textwidth,height=0.18\textheight,keepaspectratio]{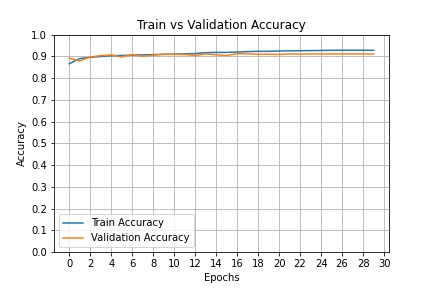}}{Diagram shows the Train versus validation for MobileNet}
			\caption{Train versus validation for MobileNet.}
			\label{fig:accuracysub1}
		\end{subfigure}
		\hfill
		\begin{subfigure}[b]{0.48\textwidth}
			\centering
			\pdftooltip{
				\includegraphics[width=\textwidth,height=0.18\textheight,keepaspectratio]{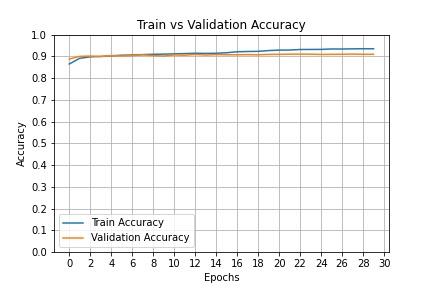}}{Diagram shows the Train versus validation for MobileNetV2}
			\caption{Train versus validation for MobileNetV2.}
			\label{fig:accuracysub2}
		\end{subfigure}
		
		\vspace{0.2cm}
		
		\begin{subfigure}[b]{0.48\textwidth}
			\centering
			\pdftooltip{
				\includegraphics[width=\textwidth,height=0.18\textheight,keepaspectratio]{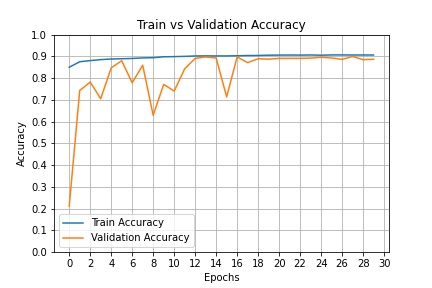}}{Diagram shows the Train versus validation for MobileNetV3}
			\caption{Train versus validation for MobileNetV3.}
			\label{fig:accuracysub3}
		\end{subfigure}
		\hfill
		\begin{subfigure}[b]{0.48\textwidth}
			\centering
			\pdftooltip{
				\includegraphics[width=\textwidth,height=0.18\textheight,keepaspectratio]{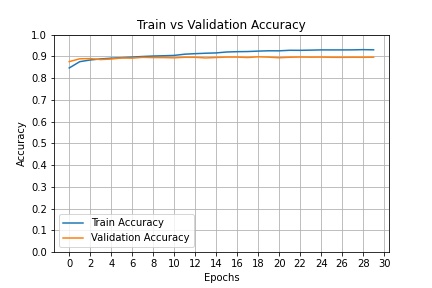}}{Diagram shows the Train versus validation for DenseNet121}
			\caption{Train versus validation for DenseNet121.}
			\label{fig:accuracysub4}
		\end{subfigure}
		
		\vspace{0.2cm}
		
		\begin{subfigure}[b]{0.48\textwidth}
			\centering
			\pdftooltip{
				\includegraphics[width=\textwidth,height=0.18\textheight,keepaspectratio]{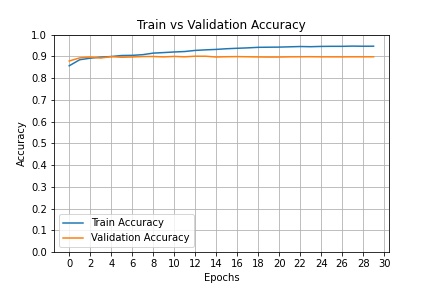}}{Diagram shows the Train versus validation for DenseNet169}
			\caption{Train versus validation for DenseNet169.}
			\label{fig:accuracysub5}
		\end{subfigure}
		\hfill
		\begin{subfigure}[b]{0.48\textwidth}
			\centering
			\pdftooltip{
				\includegraphics[width=\textwidth,height=0.18\textheight,keepaspectratio]{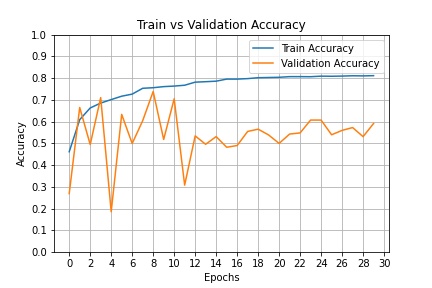}}{Diagram shows the Train versus validation for EfficientNet-B0}
			\caption{Train versus validation for EfficientNet-B0.}
			\label{fig:accuracysub6}
		\end{subfigure}
		
		\vspace{0.2cm}
		
		\begin{subfigure}[b]{0.6\textwidth}
			\centering
			\pdftooltip{
				\includegraphics[width=\textwidth,height=0.18\textheight,keepaspectratio]{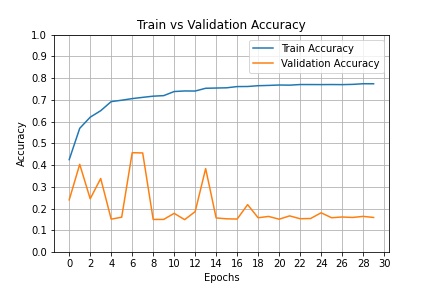}}
			{Diagram shows the Train versus validation for EfficientNet-B1}
			\caption{Train versus validation for EfficientNet-B1.}
			\label{fig:accuracysub7}
		\end{subfigure}
		
		\caption{Training and validation accuracy for the pre-trained models for detecting the DDoS attacks.}
		\label{fig:accuracy}
	\end{figure*}

	Figure~\ref{fig:accuracy} presents the training and validation accuracy curves for all seven architectures: MobileNet (Fig.~\ref{fig:accuracysub1}), MobileNetV2 (Fig.~\ref{fig:accuracysub2}), MobileNetV3 (Fig.~\ref{fig:accuracysub3}), DenseNet121 (Fig.~\ref{fig:accuracysub4}), DenseNet169 (Fig.~\ref{fig:accuracysub5}), EfficientNet-B0 (Fig.~\ref{fig:accuracysub6}), and EfficientNet-B1 (Fig.~\ref{fig:accuracysub7}). DenseNet121 and DenseNet169 show stable convergence, with training and validation accuracies increasing steadily and maintaining a minimal generalization gap throughout training. Similarly, MobileNet and MobileNetV3 exhibit consistent learning dynamics, achieving high validation accuracy with only slight divergence from their training performance, indicating strong generalization. In contrast, MobileNetV2 shows greater variability in early epochs before stabilizing at a lower validation accuracy. EfficientNet-B0 and EfficientNet-B1 exhibit noticeable fluctuations in validation accuracy across epochs, with larger gaps between training and validation performance, suggesting less stable convergence. Overall, the learning curves indicate that DenseNet-based models achieve the most stable and consistent optimization among the evaluated architectures.
	
\begin{figure*}[htbp]
		\centering
		
		\begin{subfigure}[b]{0.48\textwidth}
			\centering
			\pdftooltip{
				\includegraphics[width=\textwidth,height=0.18\textheight,keepaspectratio]{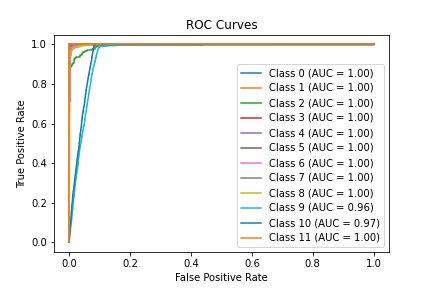}}
			{ROC results for MobileNet}
			\caption{ROC results for MobileNet.}
			\label{fig:rocsub1}
		\end{subfigure}
		\hfill
		\begin{subfigure}[b]{0.48\textwidth}
			\centering
			\pdftooltip{
				\includegraphics[width=\textwidth,height=0.18\textheight,keepaspectratio]{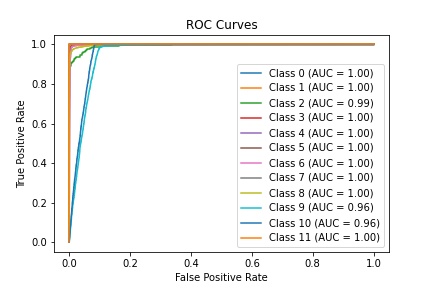}}
			{ROC results for MobileNetV2}
			\caption{ROC results for MobileNetV2.}
			\label{fig:rocsub2}
		\end{subfigure}
		
		\vspace{0.2cm}
		
		\begin{subfigure}[b]{0.48\textwidth}
			\centering
			\pdftooltip{
				\includegraphics[width=\textwidth,height=0.18\textheight,keepaspectratio]{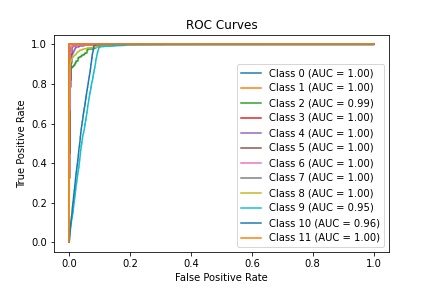}}
			{ROC results for MobileNetV3}
			\caption{ROC results for MobileNetV3.}
			\label{fig:rocsub3}
		\end{subfigure}
		\hfill
		\begin{subfigure}[b]{0.48\textwidth}
			\centering
			\pdftooltip{
				\includegraphics[width=\textwidth,height=0.18\textheight,keepaspectratio]{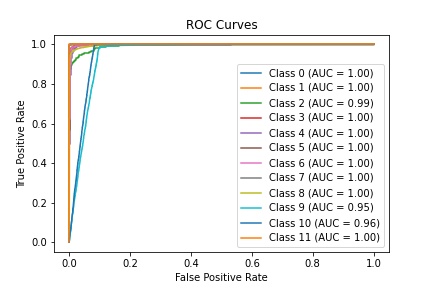}}
			{ROC results for DenseNet121}
			\caption{ROC results for DenseNet121.}
			\label{fig:rocsub4}
		\end{subfigure}
		
		\vspace{0.2cm}
		
		\begin{subfigure}[b]{0.48\textwidth}
			\centering
			\pdftooltip{
				\includegraphics[width=\textwidth,height=0.18\textheight,keepaspectratio]{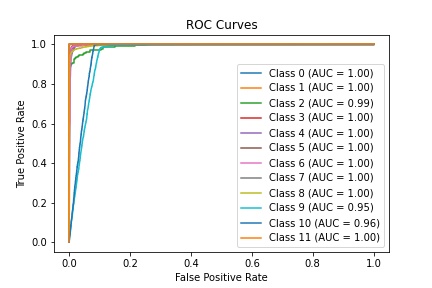}}
			{ROC results for DenseNet169}
			\caption{ROC results for DenseNet169.}
			\label{fig:rocsub5}
		\end{subfigure}
		\hfill
		\begin{subfigure}[b]{0.48\textwidth}
			\centering
			\pdftooltip{
				\includegraphics[width=\textwidth,height=0.18\textheight,keepaspectratio]{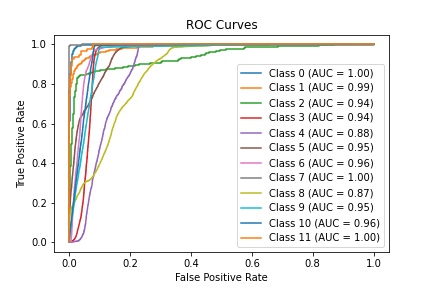}}
			{ROC results for EfficientNet-B0}
			\caption{ROC results for EfficientNet-B0.}
			\label{fig:rocsub6}
		\end{subfigure}
		
		\vspace{0.2cm}
		
		\begin{subfigure}[b]{0.6\textwidth}
			\centering
			\pdftooltip{
				\includegraphics[width=\textwidth,height=0.18\textheight,keepaspectratio]{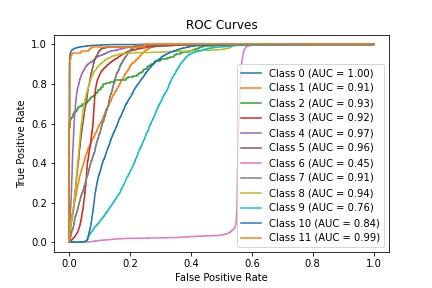}}{ROC results for EfficientNet-B1}
			\caption{ROC results for EfficientNet-B1.}
			\label{fig:rocsub7}
		\end{subfigure}
		
		\caption{The ROC diagrams for the pre-trained models for detecting the DDoS attacks.}
		\label{fig:fig_roc}
	\end{figure*}

	The aggregate AUC values reported in Table~\ref{results_compare} indicate that MobileNet achieves the highest reported AUC (0.94621), followed by MobileNetV2 (0.94364), DenseNet169 (0.94036), DenseNet121 (0.93863), and MobileNetV3 (0.93506). EfficientNet-B0 and EfficientNet-B1 achieve lower reported AUC values of 0.77878 and 0.56816, respectively. Figure~\ref{fig:fig_roc} presents the corresponding class-wise ROC curves. Differences between the aggregate results and individual class-wise curves should be interpreted in relation to the multiclass AUC aggregation method.
	
	\subsection{Statistical Reliability Evaluation}
	
	To complement conventional classification metrics, we examined two additional statistical indicators: the Matthews correlation coefficient (MCC) and Youden Index. Table~\ref{results_compare_statistics} reports these measures for the seven evaluated architectures. MCC summarizes the association between predicted and actual class labels while accounting for the multiclass confusion matrix. The Youden Index combines sensitivity and specificity to characterize class discrimination. These indicators complement accuracy, precision, recall, F1-score, AUC, and Cohen's kappa in the comparative evaluation.

	These measures are interpreted alongside classification accuracy, precision, recall, F1-score, and AUC rather than as independent evidence of deployment robustness.
	
	For multi-class classification, MCC can be expressed in terms of the confusion matrix as:
	\begin{equation}
		\mathrm{MCC}=\frac{c s-\sum_{k=1}^{K}p_k t_k}{\sqrt{\left(s^2-\sum_{k=1}^{K}p_k^2\right)\left(s^2-\sum_{k=1}^{K}t_k^2\right)}}
	\end{equation}
	\noindent
	where $c=\sum_k C_{kk}$ is the number of correct predictions, $s=\sum_{i,j}C_{ij}$ is the sample count, $p_k=\sum_i C_{ik}$ is the number predicted as class $k$, and $t_k=\sum_j C_{kj}$ is the number belonging to class $k$.
	\noindent
	The Youden index $J$ for binary classes is calculated by:
	\begin{equation}
		J = \text{Sensitivity} + \text{Specificity} - 1 = TPR + TNR - 1
	\end{equation}
	\noindent 
	where the true positive rate (sensitivity) $TPR = \frac{TP}{TP + FN}$, and the true negative rate (specificity) TNR = $\frac{TN}{TN + FP}$.
	In our multiclass task, the Youden Index $J$ is calculated by:
	\begin{equation}
		J_{macro} = \frac{1}{K} \sum_{i=1}^{K} (TPR_i + TNR_i - 1)
	\end{equation}
	\noindent
	where $K$ is the number of classes, $TPR_{i}$ is the sensitivity of class $i$, and $TNR_{i}$ is the specificity of class $i$.

	\begin{table*}[htbp]
		\centering
		\renewcommand{\arraystretch}{1.3}
		\caption{Statistical reliability comparison of pretrained models for multiclass DDoS detection.}
		\label{results_compare_statistics}
		\begin{tabular}{lcc}
			\hline
			\textbf{Model} & \textbf{MCC} & \textbf{Youden Index} \\
			\hline
			MobileNet & 0.89718 & 0.89242 \\
			MobileNetV2 & 0.89425 & 0.88728 \\
			MobileNetV3 & 0.87077 & 0.86127 \\
			DenseNet121 & 0.88212 & 0.87726 \\
			DenseNet169 & 0.88523 & 0.88071 \\
			EfficientNet-B0 & 0.55983 & 0.55756 \\
			EfficientNetB1 & 0.06944 & 0.13631 \\
			\hline
		\end{tabular}
	\end{table*}

	As shown in Table~\ref{results_compare_statistics}, MobileNet achieves the highest MCC (0.89718) and Youden Index (0.89242), followed closely by MobileNetV2, with an MCC of 0.89425 and Youden Index of 0.88728. DenseNet169 and DenseNet121 also demonstrate strong classification agreement and discrimination, whereas EfficientNet-B0 and EfficientNet-B1 exhibit substantially lower values. In particular, EfficientNet-B1 records the lowest MCC (0.06944) and Youden Index (0.13631). These findings are consistent with the classification performance results reported in Table~\ref{results_compare}.

	\section{Model Interpretability and Explainability Analysis}
	To ensure the reliability and transparency of the proposed pretrained DDoS detection models, we performed an interpretability analysis using Grad-CAM for spatial localization and SHAP for quantitative feature attribution.
	
	\subsection{Grad-CAM Spatial Attention Analysis}
	
\begin{figure*}[htbp]
		\centering
		
		\begin{subfigure}[b]{0.48\textwidth}
			\centering
			\pdftooltip{
				\includegraphics[width=\textwidth]{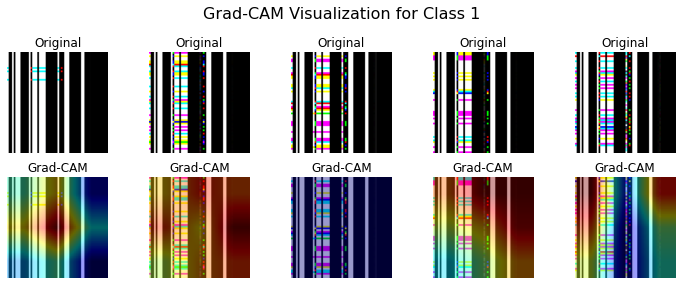}}
			{Grad-CAM results for MobileNet}
			\caption{Grad-CAM results for MobileNet.}
			\label{fig:sub1}
		\end{subfigure}
		\hfill
		\begin{subfigure}[b]{0.48\textwidth}
			\centering
			\pdftooltip{
				\includegraphics[width=\textwidth]{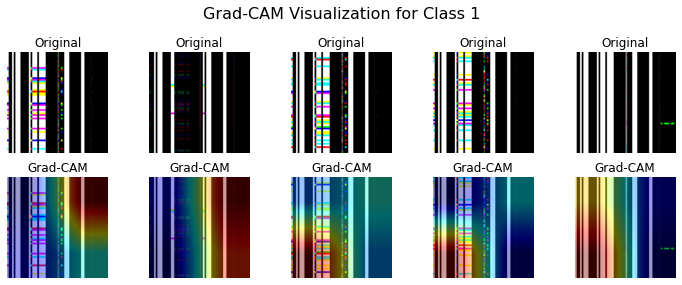}}{Grad-CAM results for MobileNetV2}
			\caption{Grad-CAM results for MobileNetV2.}
			\label{fig:sub2}
		\end{subfigure}
		
		\vspace{0.4cm}
		
		\begin{subfigure}[b]{0.48\textwidth}
			\centering
			\pdftooltip{
				\includegraphics[width=\textwidth]{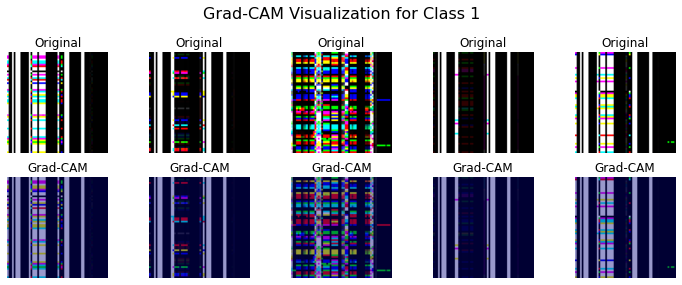}}{Grad-CAM results for MobileNetV3}
			\caption{Grad-CAM results for MobileNetV3.}
			\label{fig:sub3}
		\end{subfigure}
		\hfill
		\begin{subfigure}[b]{0.48\textwidth}
			\centering
			\pdftooltip{
				\includegraphics[width=\textwidth]{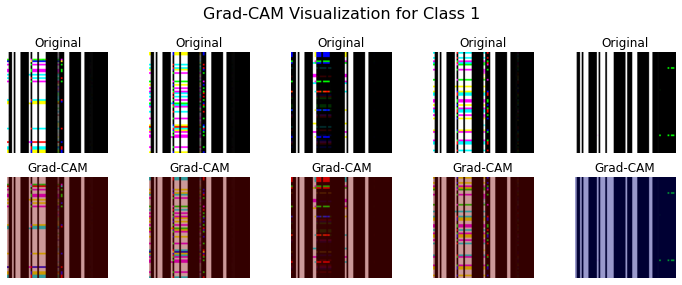}}{Grad-CAM results for DenseNet121}
			\caption{Grad-CAM results for DenseNet121.}
			\label{fig:sub4}
		\end{subfigure}
		
		\vspace{0.4cm}
		
		\begin{subfigure}[b]{0.48\textwidth}
			\centering
			\pdftooltip{
				\includegraphics[width=\textwidth]{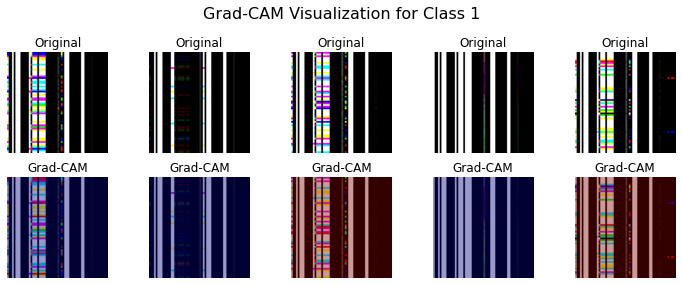}}{Grad-CAM results for DenseNet169}
			\caption{Grad-CAM results for DenseNet169.}
			\label{fig:sub5}
		\end{subfigure}
		\hfill
		\begin{subfigure}[b]{0.48\textwidth}
			\centering
			\pdftooltip{
				\includegraphics[width=\textwidth]{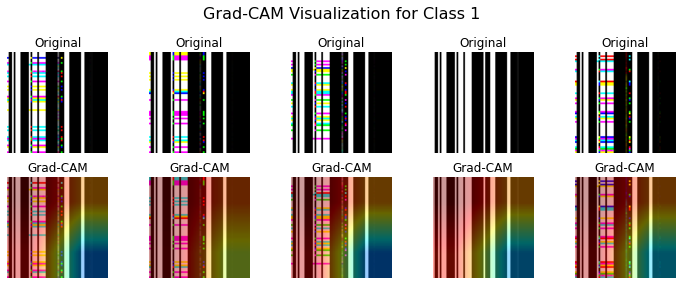}}{Grad-CAM results for EfficientNet-B0}
			\caption{Grad-CAM results for EfficientNet-B0.}
			\label{fig:sub6}
		\end{subfigure}
		
		\vspace{0.4cm}
		
		\begin{subfigure}[b]{0.6\textwidth}
			\centering
			\pdftooltip{
				\includegraphics[width=\textwidth]{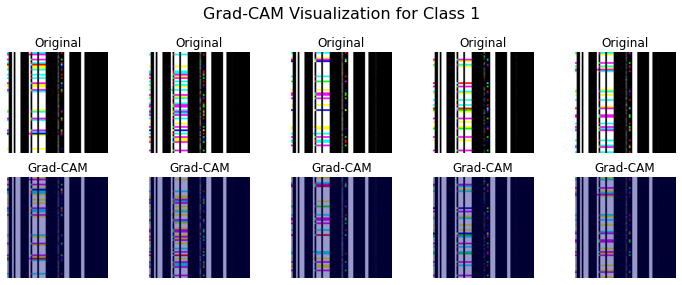}}{Grad-CAM results for EfficientNet-B1}
			\caption{Grad-CAM results for EfficientNet-B1.}
			\label{fig:sub7}
		\end{subfigure}
		
		\caption{Grad-CAM heatmaps for the pre-trained models for detecting the DDoS attacks.}
		\label{fig:gradcamTotal}
	\end{figure*}

	We generated Grad-CAM heatmaps to visualize the class-discriminative regions used by each model for all 12 traffic categories (Fig.~\ref{fig:gradcamTotal}). The top row of each figure shows the original traffic image, and the bottom row shows the corresponding Grad-CAM heatmap.

	\paragraph{\textbf{DenseNet Architectures}} DenseNet121 and DenseNet169 demonstrate highly compact and class-specific activation maps (Fig.~\ref{fig:sub4} and Fig.~\ref{fig:sub5}). Both architectures consistently highlight concentrated vertical regions within the transformed traffic representations while showing limited activation in surrounding areas. DenseNet169 exhibits slightly more compact activation patterns than DenseNet121, suggesting stronger localization of the visual features associated with its predictions.

	\paragraph{\textbf{MobileNet Family}} The Grad-CAM visualizations for the MobileNet family (Figures~\ref{fig:sub1},~\ref{fig:sub2}, and~\ref{fig:sub3}) show clear differences in spatial activation patterns across the three variants. The original MobileNet exhibits moderate localization, with activation maps highlighting concentrated regions within the transformed traffic representations, although some activation extends into adjacent areas. MobileNetV2 shows weaker activation intensities and a broader spatial distribution, indicating less concentrated localization. In contrast, MobileNetV3 exhibits more compact and consistent activation patterns, with greater concentration in localized regions and less activation in surrounding areas. Among the evaluated lightweight architectures, MobileNetV3 therefore demonstrates the most favorable qualitative spatial localization characteristics.

	\paragraph{\textbf{EfficientNet Family}} The Grad-CAM visualizations in Figures~\ref{fig:sub6} and~\ref{fig:sub7} show that the EfficientNet architectures produce smoother and more spatially diffuse activation patterns. Although the heatmaps highlight regions within the transformed traffic representations, these activations extend across broader areas rather than remaining concentrated in localized regions. This distribution suggests that the models rely on more spatially distributed visual patterns under the evaluated configurations. Among the evaluated architectures, EfficientNet-B1 exhibits the most diffuse activation patterns, indicating the lowest qualitative spatial compactness.
	
	\begin{table*}[ht]
		\centering
		\caption{Grad-CAM Spatial Localization Comparison}
		\label{tab:gradcam_comparison}
		\begin{tabular}{lcccc}
			\hline
			\textbf{Model} & \textbf{Spatial Compactness} & \textbf{Class Consistency} & \textbf{Background Suppression} & \textbf{Interpretability Level} \\
			\hline
			DenseNet169      & Very High     & Very High     & Strong         & Excellent \\
			DenseNet121      & High          & High          & Strong         & Very Good \\
			MobileNetV3      & Moderate to High & Moderate to High & Moderate       & Good \\
			MobileNet        & Moderate      & Moderate      & Moderate       & Fair--Good \\
			MobileNetV2      & Moderate      & Moderate      & Moderate to Low  & Moderate \\
			EfficientNet-B0  & Low to Moderate & Moderate      & Weak           & Moderate \\
			EfficientNet-B1  & Low           & Moderate      & Weak           & Limited \\
			\hline
		\end{tabular}
	\end{table*}
	
	Table~\ref{tab:gradcam_comparison} shows that DenseNet169 achieves the highest spatial compactness and class consistency, whereas EfficientNetB1 shows the least discriminative localization.
	
	The qualitative categories reported in Table~\ref{tab:gradcam_comparison} are defined based on visual and structural assessment of the Grad-CAM heatmaps across all twelve classes. Spatial compactness refers to how tightly activation regions concentrate around discriminative packet-density structures rather than diffusing across adjacent areas. Models categorized as Very High exhibit sharply localized vertical activation bands with minimal spread, whereas models categorized as Low exhibit broad, diffuse attention patterns. Class consistency evaluates the stability of activation behavior across different attack categories. A Very High rating indicates consistent localization patterns across all classes, while Moderate reflects variability in attention precision by attack type.
	Background suppression measures the model's ability to avoid activating non-informative regions (e.g., low-density or blank traffic areas). A Strong rating corresponds to clear suppression of irrelevant regions, whereas Weak indicates noticeable activation outside meaningful packet structures.
	These categorical assessments are derived from comparative visual inspection of heatmap compactness, intensity concentration, and suppression behavior across all models and classes. We then assign the interpretability level based on combined performance across these three criteria.

	
	\begin{figure*}[htbp]
		\centering
		
		\begin{subfigure}[b]{0.48\textwidth}
			\centering
			\includegraphics[width=\textwidth,height=0.18\textheight,keepaspectratio]{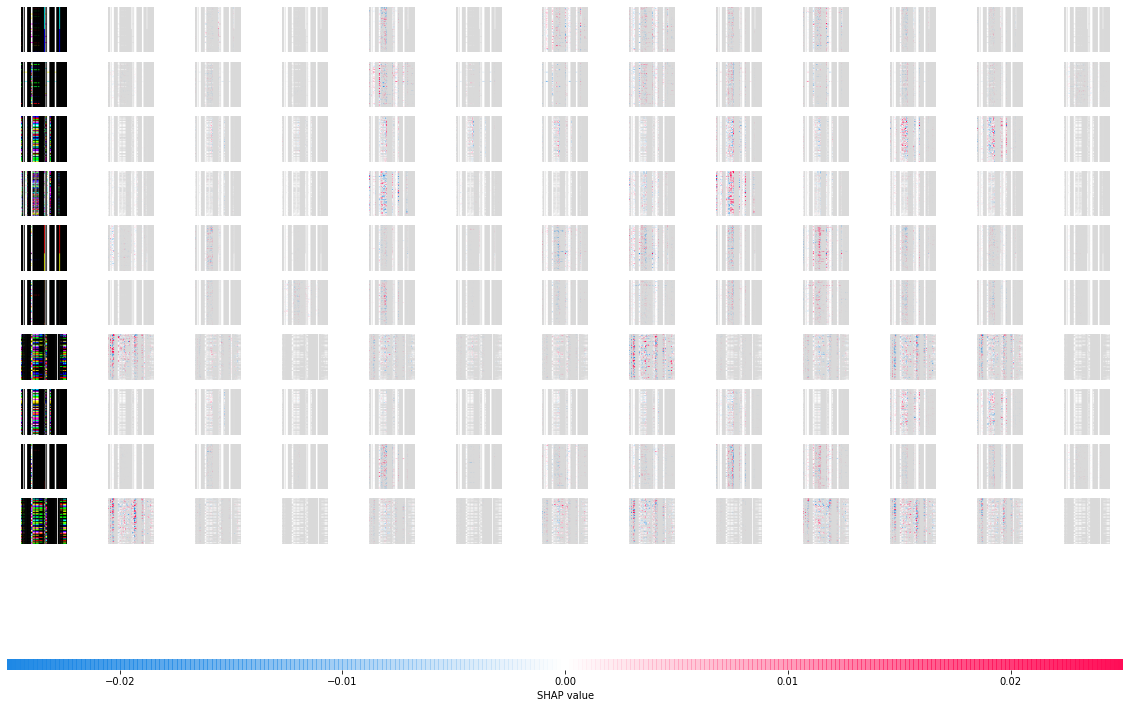}
			\caption{SHAP results for MobileNet.}
			\label{fig:shapsub1}
		\end{subfigure}
		\hfill
		\begin{subfigure}[b]{0.48\textwidth}
			\centering
			\includegraphics[width=\textwidth,height=0.18\textheight,keepaspectratio]{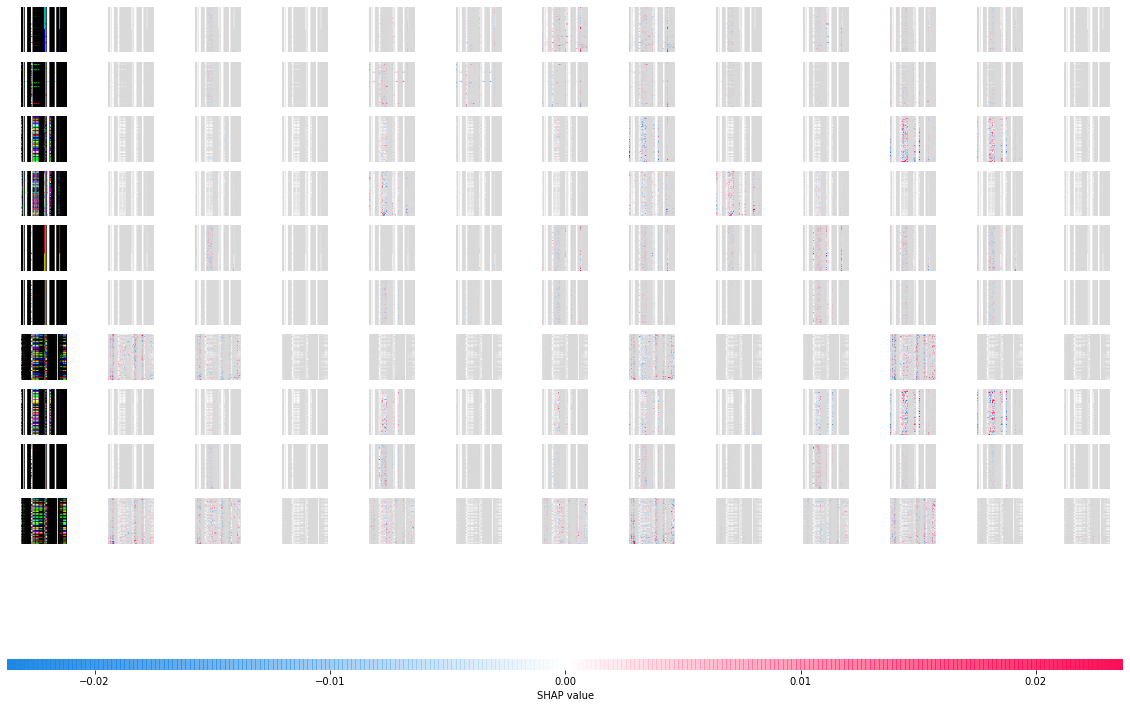}
			\caption{SHAP results for MobileNetV2.}
			\label{fig:shapsub2}
		\end{subfigure}
		
		\vspace{0.2cm}
		
		\begin{subfigure}[b]{0.48\textwidth}
			\centering
			\includegraphics[width=\textwidth,height=0.18\textheight,keepaspectratio]{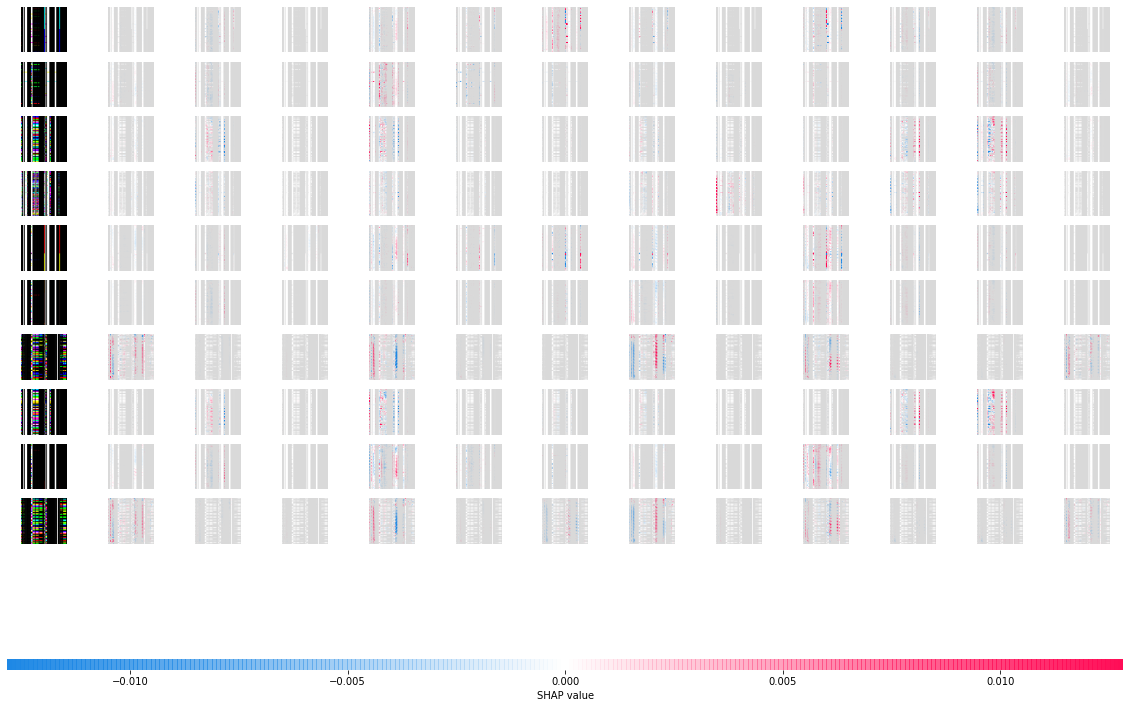}
			\caption{SHAP results for MobileNetV3.}
			\label{fig:shapsub3}
		\end{subfigure}
		\hfill
		\begin{subfigure}[b]{0.48\textwidth}
			\centering
			\includegraphics[width=\textwidth,height=0.18\textheight,keepaspectratio]{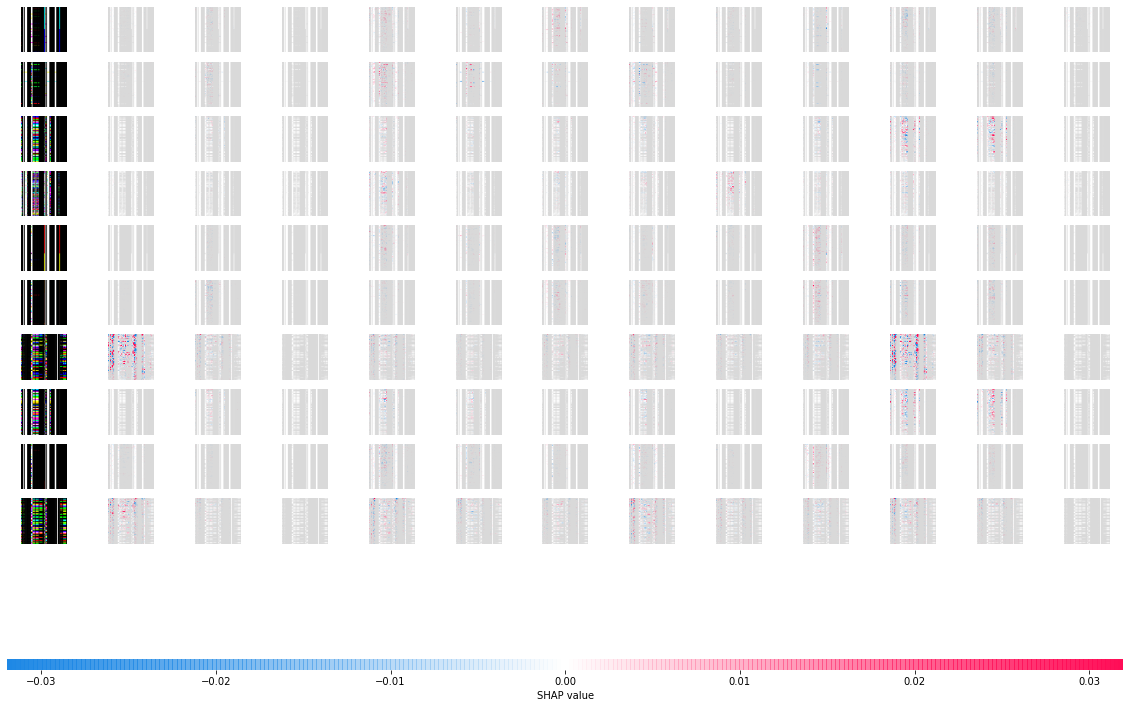}
			\caption{SHAP results for DenseNet121.}
			\label{fig:shapsub4}
		\end{subfigure}
		
		\vspace{0.2cm}
		
		\begin{subfigure}[b]{0.48\textwidth}
			\centering
			\includegraphics[width=\textwidth,height=0.18\textheight,keepaspectratio]{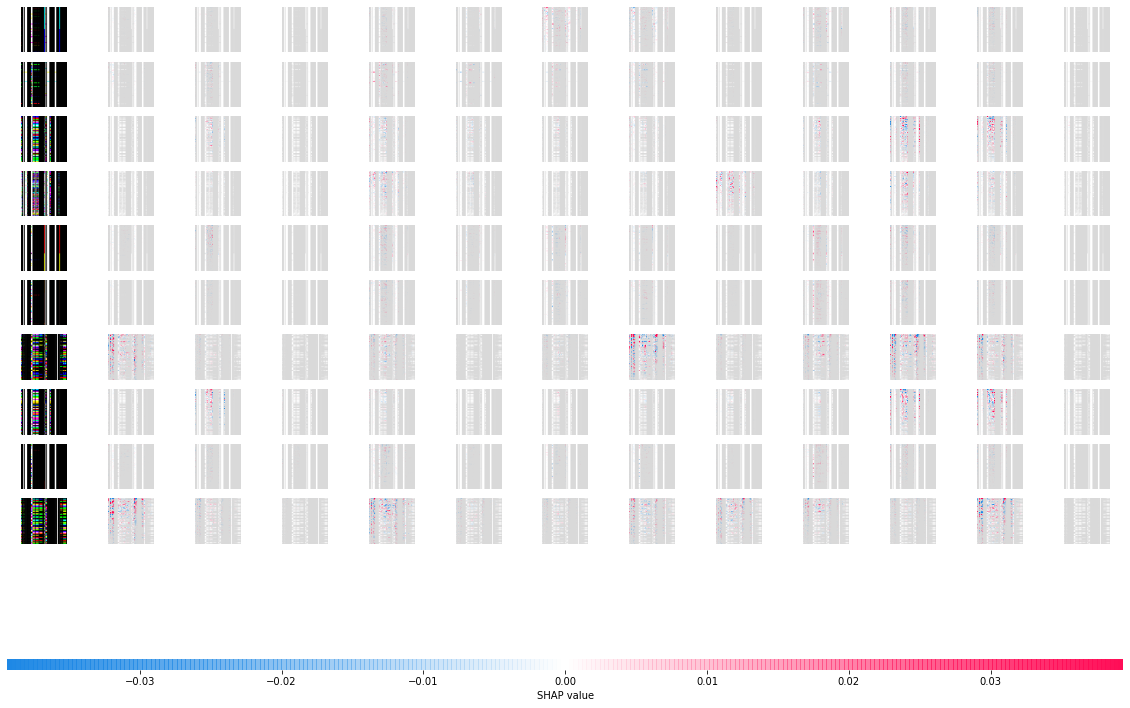}
			\caption{SHAP results for DenseNet169.}
			\label{fig:shapsub5}
		\end{subfigure}
		\hfill
		\begin{subfigure}[b]{0.48\textwidth}
			\centering
			\includegraphics[width=\textwidth,height=0.18\textheight,keepaspectratio]{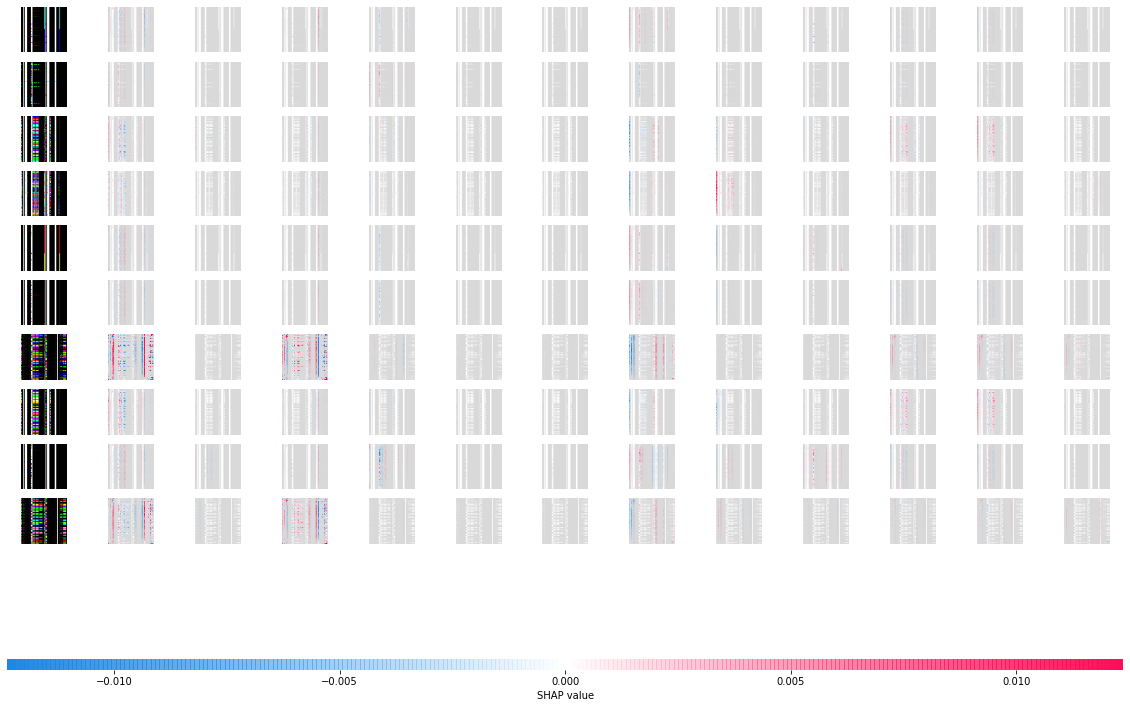}
			\caption{SHAP results for EfficientNet-B0.}
			\label{fig:shapsub6}
		\end{subfigure}
		
		\vspace{0.2cm}
		
		\begin{subfigure}[b]{0.6\textwidth}
			\centering
			\includegraphics[width=\textwidth,height=0.18\textheight,keepaspectratio]{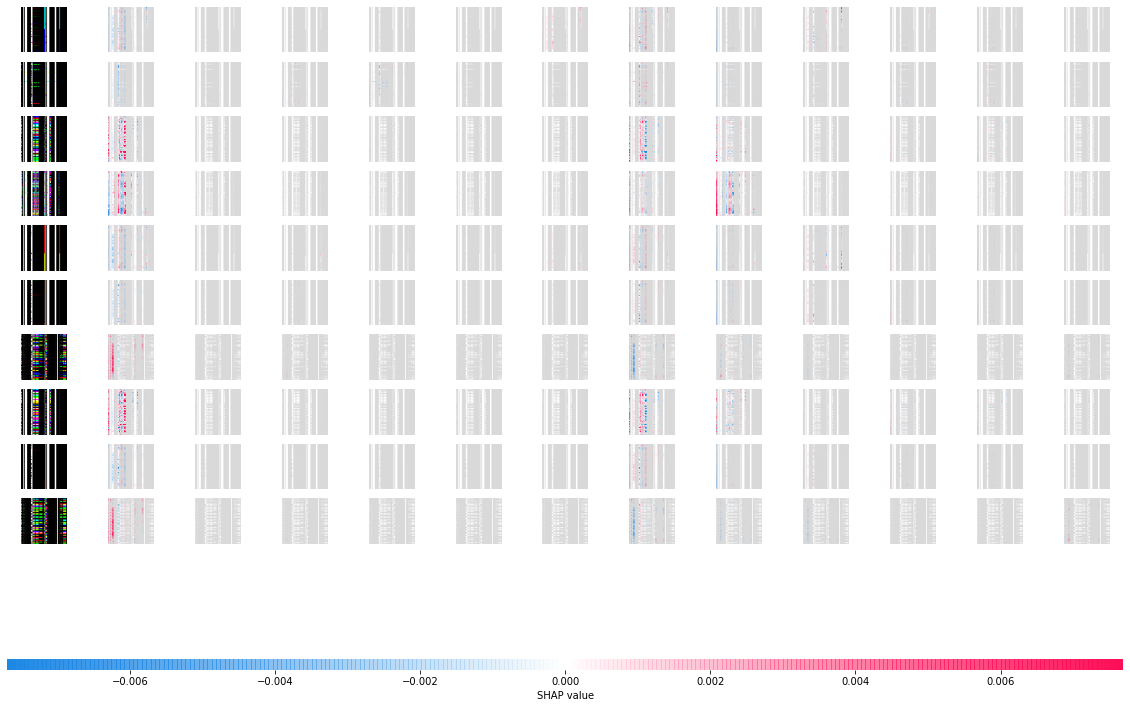}
			\caption{SHAP results for EfficientNet-B1.}
			\label{fig:shapsub7}
		\end{subfigure}
		
		\caption{The SHAP results for the pre-trained models for detecting the DDoS attacks.}
		\label{fig:shapTotal}
	\end{figure*}
	
	\subsection{SHAP Feature Attribution Analysis}
	
	We computed SHAP (SHapley Additive exPlanations) values for representative samples from each traffic class to quantitatively assess the contribution of individual spatial regions to model predictions. The main difference between Grad-CAM and SHAP is that Grad-CAM highlights regions of spatial attention. In contrast, SHAP provides a feature-level attribution score indicating the magnitude and direction of each region’s contribution to the final class prediction. Positive SHAP values correspond to features that support the predicted class. Negative SHAP values indicate suppression or counter-evidence. The SHAP visualizations are shown for MobileNet (Fig.~\ref{fig:shapsub1}), MobileNetV2 (Fig.~\ref{fig:shapsub2}), MobileNetV3 (Fig.~\ref{fig:shapsub3}), DenseNet121 (Fig.~\ref{fig:shapsub4}), DenseNet169 (Fig.~\ref{fig:shapsub5}), EfficientNet-B0 (Fig.~\ref{fig:shapsub6}), and EfficientNet-B1 (Fig.~\ref{fig:shapsub7}).
	
	\paragraph{\textbf{DenseNet Architectures}} As shown in Figure~\ref{fig:shapsub4} and Figure~\ref{fig:shapsub5}, the DenseNet architectures show the largest SHAP magnitude range, reaching approximately $\pm$0.03. The attribution maps reveal highly concentrated contribution regions within the transformed traffic images, indicating that relatively localized visual patterns contribute to the model predictions.

	These patterns indicate that the DenseNet architectures assign substantial attribution to relatively concentrated regions of the transformed traffic images. Compared with the more diffuse patterns observed in other architectures, this behavior provides clearer localization of the features contributing to the predictions. The observed attribution characteristics complement the strong classification performance reported for DenseNet121 and DenseNet169.
	
	\paragraph{\textbf{MobileNet Family}} Figures~\ref{fig:shapsub1},~\ref{fig:shapsub2}, and ~\ref{fig:shapsub3} show that MobileNet and MobileNetV3 exhibit moderate SHAP magnitudes, generally around $\pm$0.02, while MobileNetV2 displays a lower magnitude range of approximately $\pm$0.01. The attribution maps for these models are moderately compact but display some diffusion across adjacent spatial regions.
	
	MobileNetV3 demonstrates improved feature selectivity compared to MobileNetV2, with stronger concentration around discriminative packet structures and clearer suppression of irrelevant regions. In contrast, MobileNetV2 shows weaker attribution intensity and more distributed contribution patterns, indicating less decisive feature usage. Overall, the MobileNet family strikes a balance between computational efficiency and moderate interpretability, with MobileNetV3 being the most interpretable among the lightweight architectures.

	\paragraph{\textbf{EfficientNet Family}} Figures~\ref{fig:shapsub6} and~\ref{fig:shapsub7} illustrate the SHAP attribution behavior of EfficientNet-B0 and EfficientNet-B1. EfficientNet-B0 shows moderate magnitude contributions (approximately $\pm$0.02), but the attribution patterns are noticeably more diffuse than in the DenseNet models. Although relevant regions are activated, contributions are distributed across broader spatial areas, suggesting smoother evidence aggregation rather than sharp discriminative emphasis.
	
	EfficientNet-B1 demonstrates the smallest SHAP magnitude range (approximately $\pm$0.006) among the evaluated models. Its attribution maps show highly diffuse contributions with limited strong positive or negative peaks. Compared with the more concentrated patterns observed for DenseNet, this distribution provides less clearly localized visual evidence for the model's predictions.
	
	Table~\ref{tab:shap_comparison} shows quantitative SHAP attribution characteristics across all evaluated architectures.
	
	\begin{table*}[ht]
		\centering
		\caption{SHAP Attribution Comparison Across Architectures}
		\label{tab:shap_comparison}
		\begin{tabular}{llp{2.5cm}lp{2cm}}
			\hline
			\textbf{Model} & \textbf{SHAP Magnitude} & \textbf{Attribution Concentration} & \textbf{Negative Suppression} & \textbf{Attribution Strength} \\
			\hline
			DenseNet169     & $\pm$0.03      & Very Compact     & Strong          & Very Strong \\
			DenseNet121     & $\pm$0.03      & Compact          & Strong          & Strong \\
			MobileNetV3     & $\pm$0.02--0.025 & Moderate        & Moderate        & Moderate--Strong \\
			MobileNet       & $\pm$0.02      & Moderate         & Moderate        & Moderate \\
			EfficientNet-B0 & $\pm$0.02      & Diffused         & Moderate--Weak  & Moderate \\
			MobileNetV2     & $\pm$0.01      & Less Compact     & Moderate--Weak  & Weak--Moderate \\
			EfficientNet-B1 & $\pm$0.006     & Highly Diffused  & Weak            & Weak \\
			\hline
		\end{tabular}
	\end{table*}

	\subsection{Integrated Interpretability Comparison}

	The Grad-CAM and SHAP analyses reveal consistent qualitative patterns across the evaluated architectures. Models exhibiting concentrated Grad-CAM activation regions generally also show more localized SHAP contributions, whereas architectures with diffuse activation patterns tend to exhibit more distributed feature attributions. This agreement across two complementary explanation methods supports the comparative interpretation of the observed attribution patterns.
	
	DenseNet169 consistently achieves the highest spatial compactness, the strongest SHAP attribution magnitude, and the most effective background suppression. DenseNet121 follows closely, demonstrating similarly compact but slightly less intense attribution behavior. MobileNetV3 is the strongest lightweight architecture in terms of interpretability, while EfficientNet-B1 shows the least discriminative localization and feature decisiveness.
	
	The interpretability findings complement the classification and statistical results by revealing differences in how the evaluated architectures distribute attribution across the transformed traffic representations. These observations provide an additional dimension for model comparison beyond predictive performance and computational cost.
	
	Table~\ref{tab:overall_interpretability} summarizes the overall interpretability ranking derived from the integrated analysis.
	
	\begin{table}[ht]
		\centering
		\caption{Overall Interpretability Ranking Based on Grad-CAM and SHAP Analysis}
		\label{tab:overall_interpretability}
		\begin{tabular}{lc}
			\hline
			\textbf{Model} &\textbf{Rank}  \\
			\hline
			DenseNet169 &1 \\
			DenseNet121 &2\\
			MobileNetV3 &3\\
			MobileNet &4\\
			EfficientNet-B0&5 \\
			MobileNetV2 &6\\
			EfficientNet-B1 &7\\
			\hline
		\end{tabular}
	\end{table}

	Interpretability plays an important role in intrusion detection systems, particularly where false positives and false negatives can have significant operational consequences. Grad-CAM and SHAP provide complementary views of the regions and features associated with model predictions, allowing differences in attribution patterns to be examined across architectures.
	
	The DenseNet architectures exhibit concentrated Grad-CAM activation patterns and comparatively strong SHAP attributions. These observations provide more clearly localized visual explanations than the diffuse attribution patterns observed for some other architectures. Such characteristics may assist analysts in examining detection decisions, although explanation fidelity and operational usefulness require separate validation.
	
	Accordingly, the results identify DenseNet architectures as promising candidates when qualitative attribution characteristics are prioritized, while deployment decisions must also consider predictive performance, computational cost, and validation in the intended operational environment.
	
	\section{Discussion}
	In this section, we synthesize the empirical findings across detection performance, computational cost, reliability metrics, and model interpretability, with emphasis on practical deployment for IoT-driven systems and fog-level intrusion detection. The evaluation spans three architecture families: MobileNet, DenseNet, and EfficientNet. It integrates conventional performance metrics (Table~\ref{results_compare}), computational cost indicators (Table~\ref{tab:time_cost}), reliability-oriented statistics (Table~\ref{results_compare_statistics}), and explainability evidence from Grad-CAM and SHAP analyses (Figures~\ref{fig:gradcamTotal} and~\ref{fig:shapTotal}, Tables~\ref{tab:gradcam_comparison} and~\ref{tab:shap_comparison}).
	
	\subsection{Performance Trade-offs Across Architectures}
	MobileNet, DenseNet121, and DenseNet169 achieve consistently high overall accuracy (approximately 98\%) and strong reported classification performance across the evaluated architectures (Table~\ref{results_compare}). Among these, MobileNet achieves the highest reported accuracy (98.476\%), while DenseNet169 maintains comparable accuracy (98.298\%) with strong agreement statistics (Kappa $\approx$ 0.885) and robust correlation-based reliability (MCC $\approx$ 0.885; Table~\ref{results_compare_statistics}). These results show that DenseNet connectivity offers a favorable inductive bias for exploiting repeated packet-structure patterns introduced by the tabular-to-image transformation (Figure~\ref{Data_Transformation}).

	In contrast, the EfficientNet architectures exhibit weaker classification performance under the evaluated configuration, particularly EfficientNet-B1, which achieves an accuracy of 85.998\%, Cohen's kappa of 0.05826, and MCC of 0.06944. These results indicate substantially weaker agreement and class discrimination compared with the MobileNet and DenseNet architectures. This indicates that despite EfficientNet’s reputation for accuracy-efficiency trade-offs in natural image recognition, it does not transfer robustly to the highly structured, synthetic-like texture patterns represented in network traffic images under the current fine-tuning scheme. EfficientNet-B0 performs better than B1 but remains substantially lower compared to MobileNet and DenseNet models, with reduced precision/recall and lower discriminative capacity (AUC $\approx$ 0.779; Table~\ref{results_compare}).
	
	\subsection{ROC Behavior and Class-wise Separability}
	The aggregate AUC results in Table~\ref{results_compare} show relatively similar discrimination among the MobileNet and DenseNet architectures, ranging from 0.93506 for MobileNetV3 to 0.94621 for MobileNet. MobileNetV2 achieves an AUC of 0.94364, exceeding MobileNetV3 (0.93506), despite differences in their other performance metrics. DenseNet169 and DenseNet121 achieve AUC values of 0.94036 and 0.93863, respectively.
	
	In contrast, EfficientNet-B0 and EfficientNet-B1 obtain substantially lower reported aggregate AUC values of 0.77878 and 0.56816, respectively. These results indicate weaker aggregate class discrimination under the evaluated configurations. The class-wise ROC curves in Figure~\ref{fig:fig_roc} provide additional visual information about variation across traffic categories; however, class-specific conclusions require examination of the corresponding individual AUC values.
	
	\subsection{Training Dynamics and Generalization Stability}
	The train-versus-validation learning curves (Figure~\ref{fig:accuracy}) provide additional evidence about optimization stability and generalization. DenseNet121 and DenseNet169 show smooth convergence and a small, stable generalization gap across epochs. This indicates consistent optimization and robust generalization under the selected fine-tuning strategy. MobileNet and MobileNetV3 also converge steadily with high validation accuracy and limited divergence from training behavior, meaning that these architectures generalize effectively while maintaining computational efficiency.

	On the other hand, EfficientNet-B0 and EfficientNet-B1 show noticeable validation oscillations and larger train--validation gaps. This instability aligns with their lower performance and reliability statistics (Tables~\ref{results_compare} and~\ref{results_compare_statistics}) and indicates sensitivity to hyperparameters, fine-tuning depth, or a mismatch between pretraining priors and the target domain. MobileNetV2 also shows higher early-epoch variability before stabilization, which may partially explain its weaker discriminative and interpretability behavior relative to MobileNetV3.
	
	\subsection{Cost and Efficiency for Fog-level Deployment}
	From a deployment perspective, the cost evaluation demonstrates a clear trade-off between latency and accuracy across the model families. MobileNet achieves the lowest inference time per sample (0.000530\,s), followed by MobileNetV2 and MobileNetV3 (Table~\ref{tab:time_cost}). This makes MobileNet variants highly attractive for resource-constrained IoT/fog implementations. DenseNet models, while robust and explainable, show higher inference time (DenseNet169: 0.003050\,s). This reflects the computational overhead of deeper dense connectivity. EfficientNet-B0 and EfficientNet-B1 fall between MobileNet and DenseNet in inference time. However, their performance and reliability limitations reduce their practical suitability under strict implementation constraints.
	
	Training time also varies substantially, with MobileNet requiring roughly 62 minutes and DenseNet169 requiring approximately 346 minutes (Table~\ref{tab:time_cost}). While training costs may be diluted in many deployment scenarios, they matter more in scenarios with frequent model updates, online adaptation, or edge implementation. Therefore, MobileNet is the strongest candidate when prioritizing the measured inference latency and classification accuracy, achieving the lowest latency (0.000530\,s per sample) and the highest accuracy (98.476\%). MobileNetV3 may be considered when its qualitative interpretability characteristics are given additional weight, while DenseNet169 offers the strongest qualitative attribution characteristics among the evaluated architectures, although at a higher inference cost. These deployment implications remain provisional because the timing measurements were obtained on the experimental hardware rather than a dedicated fog device.
	
	\subsection{Reliability Metrics and Decision Confidence}
	These reliability gaps are important in real-time IoT environments because risk is not determined solely by average accuracy. The reliability-oriented statistics in Table~\ref{results_compare_statistics} strengthen the interpretation of the performance results. Models with high MCC and Youden Index provide more trustworthy multi-class decision behavior, especially under uneven class distributions. DenseNet169 and MobileNet achieve strong MCC values near 0.885--0.897 and high Youden Index values near 0.88--0.89, supporting their suitability for operational deployment. On the other hand, EfficientNet-B1 shows extremely low MCC and Youden Index values, indicating unreliable class discrimination and weak sensitivity--specificity balance under the assessed conditions.

	\subsection{Explainability: Agreement Between Grad-CAM and SHAP}
	This paper's main contribution is an explainability-aware evaluation of transfer learning architectures for DDoS detection. The Grad-CAM results (Figure~\ref{fig:gradcamTotal} and Table~\ref{tab:gradcam_comparison}) show that DenseNet architectures produce the most compact and class-consistent spatial localization patterns, with strong background suppression. MobileNetV3 shows greater compactness compared to MobileNetV2 and the base MobileNet, supporting its role as the most interpretable lightweight model. EfficientNet architectures, especially EfficientNet-B1, show the most diffuse attention with weaker suppression of non-informative regions, indicating less discriminative localization.
	
	The SHAP analysis (Figure~\ref{fig:shapTotal} and Table~\ref{tab:shap_comparison}) quantitatively demonstrates these trends. DenseNet models achieve the highest attribution magnitude (approximately $\pm$0.03) with compact contributions and clear positive--negative separation, reflecting decisive feature reliance and structured evidence. MobileNetV3 exhibits moderate-to-strong attribution magnitude (approximately $\pm$0.02--0.025) with improved selectivity relative to MobileNetV2. EfficientNet-B1 shows the smallest magnitude (approximately $\pm$0.006) and highly diffused attribution, indicating weaker decisiveness and less transparent decision logic.
	
	The qualitative agreement between Grad-CAM spatial localization and SHAP feature attribution supports the comparative interpretation of the observed explanation patterns. Architectures with concentrated Grad-CAM activations generally exhibit more localized SHAP contributions, consistent with the integrated interpretability ranking in Table~\ref{tab:overall_interpretability}. This complementary evidence is useful for comparing the models' attribution characteristics, although additional evaluation would be required to establish explanation fidelity and operational trustworthiness.
	
	\subsection{Implications for IoT-driven Business Systems}
	For IoT-driven business systems, the choice of model depends on the balance between operational constraints and trust requirements. Lightweight MobileNet models minimize latency and suit fog/edge implementation where rapid filtering and early warning are required. Within this family, MobileNetV3 offers a strong compromise, providing high performance and more coherent explanations than MobileNetV2. DenseNet models, particularly DenseNet169, provide the strongest evidence of interpretability and reliability, making them suitable for scenarios where decisions must be auditable and defensible (e.g., regulated infrastructures, critical services, or environments requiring forensic analysis).
	EfficientNet architectures are less favorable in the evaluated configuration due to reduced stability, weaker reliability, and lower interpretability clarity. This does not imply that EfficientNet is inherently unsuitable for cybersecurity, but rather that the architectural priors and fine-tuning configuration may require further adaptation for traffic-image representations (e.g., alternative unfreezing strategies, stronger regularization, or architecture-specific preprocessing).
	
	Thus, the results support a deployment-oriented interpretation: MobileNet is the strongest candidate when prioritizing measured inference latency and classification accuracy, whereas MobileNetV3 may be considered when qualitative interpretability is assigned greater importance. DenseNet169 offers the most favorable qualitative attribution characteristics among the evaluated architectures, although its higher computational cost may constrain deployment on resource-limited hardware.

	\section{Security Robustness and Limitations}
	
	While the proposed evaluation shows strong detection performance across multiple DDoS categories, several security-related limitations warrant consideration.
	
	First, the evaluation uses the CICDDoS2019 benchmark dataset. Although comprehensive, real-world deployments may encounter unseen attack patterns, distribution shifts, or zero-day attack variants that could impact model generalization.
	
	Second, this study assumes a closed-world setting where attack categories are known during training. The detection capability against novel or adversarially modified attacks remains an open research challenge.
	
	Third, the framework does not explicitly evaluate adversarial evasion techniques such as traffic perturbation or adversarial feature manipulation. Future work may explore adversarial robustness and online adaptive learning mechanisms.
	
	Finally, encrypted traffic environments and payload obfuscation may alter feature distributions, potentially affecting detection reliability.
	
	Despite these limitations, the presented evaluation provides a structured comparison of pretrained models using classification performance, computational cost, statistical indicators, and explainability analyses relevant to resource-constrained IoT deployment.
	
	\section{Conclusion}
	This paper proposed a comprehensive empirical evaluation of transfer learning architectures for DDoS detection in IoT-driven business systems, emphasizing not only predictive performance but also computational cost, reliability, and explainability. Using the CICDDoS2019 benchmark and an image-based transformation of network traffic, we fine-tuned and compared seven pretrained models under a unified experimental protocol.
	
	The results demonstrate that MobileNet- and DenseNet-based architectures achieve strong multi-class detection performance (approximately 98\% accuracy), with high agreement and reliability indicators, including Cohen’s Kappa, MCC, and Youden Index. MobileNet achieves the highest reported classification accuracy (98.476\%) and the strongest MCC (0.89718), Cohen's kappa (0.89711), and Youden Index (0.89242) among the evaluated architectures. DenseNet169 achieves comparable classification accuracy (98.298\%) and exhibits the most favorable qualitative Grad-CAM and SHAP attribution characteristics. MobileNetV3 provides an alternative with relatively low inference latency and more concentrated qualitative attribution patterns than MobileNetV2. EfficientNet-B0 and EfficientNet-B1 exhibit weaker aggregate classification performance and reliability statistics under the evaluated configurations.
	
	From a practical deployment perspective, the findings highlight trade-offs among predictive performance, computational cost, and qualitative interpretability. MobileNet achieves the lowest measured inference latency and highest classification accuracy, making it a promising candidate for latency-sensitive applications. DenseNet169 exhibits the strongest qualitative interpretability characteristics, while MobileNetV3 provides an alternative when attribution compactness is prioritized alongside relatively low inference latency. These findings require validation on representative fog or edge hardware before operational deployment claims can be established.
	
	Future work will extend this study by incorporating deeper, confusion-matrix-driven, class-specific error analysis; evaluating robustness under distribution shifts (e.g., unseen attack variants); and exploring edge-aware compression and quantization strategies to reduce inference cost further while maintaining classification performance and explainability characteristics.

	\section*{Data Availability Statement}
	This paper uses the publicly available CICDDoS2019 dataset, which contains labeled network traffic representing benign activity and multiple distributed denial-of-service (DDoS) attack categories. The dataset is accessible at https://www.unb.ca/cic/datasets/ddos-2019.html. No new data were collected for this study.
	
	\section*{Declaration of Generative AI}
	During the preparation of this manuscript, the authors used Grammarly, ChatGPT, and Gemini to assist with language editing, grammatical corrections, and LaTeX formatting. All AI-assisted suggestions were reviewed and revised by the authors, who take full responsibility for the scientific content, accuracy, originality, and integrity of the manuscript.
	
	\section*{Disclosure statement}
	The authors report there are no competing interests to declare.
	
	%

	


	\ifCLASSOPTIONcaptionsoff
	\newpage
	\fi

	\bibliographystyle{IEEEtran}
	\bibliography{references}
	
\end{document}